# Exploring multi-dimensional spaces: a Comparison of Latin Hypercube and Quasi Monte Carlo Sampling Techniques

*Sergei Kucherenko[1*], Daniel Albrecht[2], Andrea Saltelli[3]*

[1]*CPSE, Imperial College London, London, SW7 2AZ, UK*
[2]*The European Commission, Joint Research Centre,
TP 361, 21027 ISPRA (VA), Italy*
[3]*European Centre for Governance in Complexity,
Universitat Autonoma de Barcelona, Catalonia, Spain*

**Abstract**

Three sampling methods are compared for efficiency on a number of test problems of various complexity for which analytic quadratures are available. The methods compared are Monte Carlo with pseudo-random numbers, Latin Hypercube Sampling, and Quasi Monte Carlo with sampling based on Sobol' sequences. Generally results show superior performance of the Quasi Monte Carlo approach based on Sobol' sequences in line with theoretical predictions. Latin Hypercube Sampling can be more efficient than both Monte Carlo method and Quasi Monte Carlo method but the latter inequality holds for a reduced set of function typology and at small number of sampled points. In conclusion Quasi Monte Carlo method would appear the safest bet when integrating functions of unknown typology.

## 1. Introduction

In many practical applications numerical models are used in a parametric bootstrap fashion as to obtain a distribution of the inference, from which e.g. some estimate of model statistics (average, percentiles) can be obtained. This implies propagating through the model – assumed deterministic here - the uncertainty from a set of the model's input variables or factors. Since models can be expensive to run, it is sometimes impossible to have as many model runs as needed to achieve the desired accuracy. In such cases an efficiency of the Monte Carlo sampling technique can make the difference between a reliable inference and a misleading result.

The simple Monte Carlo technique is known to have low efficiency. Variance reduction methods can be employed to increase its efficiency. One of the most popular variance reduction technique is stratified sampling using Latin Hypercube Sampling (LHS) [1]. Many studies have been made over the years to develop LHS with better space filling properties. Some authors have proposed to improve LHS space filling not only in one dimensional projection, but also in higher dimensions [2]. Different optimality criterion such as entropy, integrated mean square error, minimax and maximin distances, and others can be used for optimizing LHS [3]. The maximin criterion, consisting in maximizing the minimal distance between points to avoid sampling designs with points too close to one another, was used in [4]. Iooss et al [5] compared the performance of different types of space filling designs, in the class of the optimal LHS, in terms of the efficiency of the subsequent metamodel fitting. They concluded that optimized LHS with minimal wrap-around discrepancy (defined later in this work) are particularly well-suited for the Gaussian process metamodel fitting.

\* Corresponding author at: Imperial College London, London, SW7 2AZ, UK.
E-mail address: s.kucherenko@imperial.ac.uk (S. Kucherenko).

A detailed study on LHS and various modified designs of LHS was given by Owen [6], [7], [8], Tang, [9], Kai-Tai Fang and co-authors [10-11] and some other authors.

Kalagnanam and Diwekar [12] compared the performance of the Latin hypercube and Hammersley sampling techniques on a number of test problems. The number of samples required for convergence to mean and variance was used as a measure of performance. It was shown that the sampling technique based on the low discrepancy Hammersley points requires far fewer samples to converge to the variance of the tested distributions: for the selected test cases Hammersley required up to 40 times fewer sampled points than LHS.

Wang et al. [13] developed a new sampling technique combining LHS and Hammersley sequence sampling and compared it with MC, LHS and the Hammersley sequence sampling methods for calculating uncertainty in risk analysis. They found that the new (LHS+Hammersley) sampling technique and the Hammersley sequence sampling technique behaved consistently better than MC and pure LHS.

While its application in areas like experimental design and metamodel fitting is well studied, the efficiency of LHS in other areas such as high dimensional integration has not been analyzed systematically. In high dimensional integration, efficiency depends upon the degree of uniformity of the sampled points in the *n*-dimensional space. In this work we compare efficiencies of three sampling methods: the MC method with pseudo-random, LHS sampling and the QMC method with sampling based on Sobol' sequences. Although we discuss three different LHS designs, only un-optimized LHS is used in our numerical experiment. In spite of the recent advances in optimization of high dimensional properties of LHS, due to the NP complexity of the optimization, these advances are only practical for relatively small designs. The most efficient and hence expensive optimized LHS designs cannot be used for high dimensional integration. Furthermore we note a widespread practice of use of non-optimized LHS which justifies the present exercise.

QMC sampling is performed by using the high-dimensional Sobol' Low Discrepancy Sequence (LDS) generator with advanced uniformity properties: Property A for all dimensions and Property A' for adjacent dimensions (see Section 5 for details). Three different classes of test functions are considered for high dimensional integration, where the classification of functions is based on their effective dimension as discussed in [14]. We report the absolute values of integration errors as well as the convergence rate.

This paper is organized as follows: in the next Section we describe Monte Carlo integration method. Section 3 contains a brief description of LHS. The QMC method is presented in Section 4. Section 5 introduces Sobol' Low Discrepancy Sequences (LDS). A first didactic set of results comparing different techniques for populating a unit square is given in Section 6. Discrepancy as a quantitative measure for the deviation of sampled points from the ideal (desired) uniform distribution is described in Section 7, where $L_2$ discrepancy for the three considered sampling methods is also reported. Global sensitivity analysis and effective dimensions are introduced in Section 8. The results from the three sampling techniques are compared and discussed in Sections 9 and 10. Section 11 concludes our analysis.

## 2. Monte Carlo integration

Consider the integral of the square-integrable function $f(x)$ over the $n$-dimensional unit hypercube $H^n$

$$I[f] = \int_{H^n} f(x)dx. \qquad (2.1)$$

Function $f(x)$ is assumed to be integrable in $H^n$.

The MC quadrature formula is based on the probabilistic interpretation of an integral. For a random variable $x$ that is uniformly distributed in $H^n$

$$I[f] = E[f(x)],$$

where $E[f(x)]$ is the mathematical expectation. An approximation to this expectation is

$$I_N[f] = \frac{1}{N}\sum_{i=1}^{N} f(x_i), \qquad (2.2)$$

where $x_i = (x_i^1,...,x_i^n)$, $i=1,...,N$ is a sequence of independent random points in $H^n$ of length $N$. In other words, $N$ is a number of sampled points. Approximation (2.2) is known as the simple (or crude) Monte Carlo estimate of the integral. The approximation $I_N[f]$ converges to $I[f]$ with probability 1.

It follows from the Central Limit Theorem that the expectation of an integration error $\varepsilon^2$, where $\varepsilon = |I[f] - I_N[f]|$ is

$$E(\varepsilon^2) = \frac{\sigma^2(f)}{N},$$

where $\sigma^2(f)$ is the variance given by

$$\sigma^2(f) = \int_{H^n} f^2(x)dx - (\int_{H^n} f(x)dx)^2 .$$

The expression for the root mean square error of the MC method is

$$\varepsilon_N = (E(\varepsilon^2))^{1/2} = \frac{\sigma(f)}{N^{1/2}}. \qquad (2.3)$$

The convergence rate of MC does not depend on the number of variables $n$ although it is rather low. Various variance reduction techniques (*e.g.* antithetic method, control variates, stratified sampling, importance sampling) can be applied to reduce the value of the numerator in (2.3), which does not change the MC integration convergence rate of $O(1/N^{1/2})$. These techniques are not considered here.

The efficiency of MC methods is determined by the properties of random numbers. It is known that random number sampling is prone to clustering: for any sampling there are always empty areas as well as regions in which random points are wasted due to clustering; as new points are added randomly, they do not necessarily fill the gaps between already sampled points. The quality of a quadrature by a finite number of sampled points depends on the uniformity of the points' distribution, not their randomness. Sampling strategies aimed at placing points more uniformly include the LHS and LDS (also known as quasi random) designs addressed in the present analysis.

### 3. Latin Hypercube Sampling

LHS is widely applied in computational engineering. It was developed by McKay et al. [1]. It is one form of stratified sampling that can reduce the variance in the Monte Carlo estimate of the integrand. It was further analyzed by Iman and Shortencarier [15] and then by Stein [16]. It has been shown that LHS can improve the efficiency compared to the Monte Carlo approach, though we show here (Section 9) that this only holds for certain classes of functions.

Consider the range [0,1] divided into $N$ intervals of the equal length $1/N$. One point is selected at random from each interval forming a sequence of $N$ points in $H^1$ $\{x_i^1\}$, $i = 1,..,N$. Similarly but independently we construct another sequence $\{x_i^2\}$, $i = 1,..,N$. The two sequences $\{x_i^1\}$, $i = 1,..,N$ and $\{x_i^2\}$, $i = 1,..,N$ can be paired to populate a bidimensional space. These $N$ pairs can in turn randomly be combined with the $N$ values of $\{x_i^3\}$, $i = 1,...,N$ to form $N$ triplets, and so on until an $n$-dimensional sequence of $N$ is formed.

The algorithm formally can be presented as follows. Let $\{\pi_k\}$, $k = 1,...,n$ be independent random permutations of $\{1,...,N\}$ each uniformly distributed over all $N!$ possible permutations. Set

$$x_i^k = \frac{\pi_k(i) - 1 + U_i^k}{N}, \ i = 1,...,N, \ k = 1,...,n, \qquad (3.1)$$

where $U_i^k$ are independent randomly sampled points on $[0,1]$ interval.

It is easy to see that only one point of $\{x_i^k\}$, $i = 1,...,N$, $k = 1,...,n$ falls between $(i-1)/N$ and $i/N$, $i = 1,..,N$ for each dimension $k = 1,..,n$. However, this stratification scheme is built by superimposing well stratified one-dimensional samples, and cannot be expected to provide in principle good uniformity properties in a $n$-dimensional unit hypercube $H^n$.

In this work we used the code for the LHS from [17], referred to as "standard LHS" in the following.

The efficiency of the standard LHS can be improved by taking the standard LHS as a starting design and then optimizing it according to some optimization criterion: e.g. maximizing the minimum distance between any two points (maxmin criterion), minimizing the distance between a point of the input domain and the points of the design or minimizing the discrepancy. Optimization can be done by using different methods: choice of the best (in terms of the chosen criteria) LHS amongst a large number of different LHS, column wise pair wise exchange algorithms or optimization techniques such as genetic algorithms, simulated annealing, and others [11]. It has been shown by Iooss *et al.* [5] that minimizing the discrepancy leads to a better space-filling design compared to the one where the minimum distance is maximized. In particular they compared the two-dimensional projections of the maximin LHS and low wrap-around discrepancy LHS with $N = 100$ points and different initial dimensions. They found that the initial LHS design optimized with the wrap-around discrepancy offers the best results. These results were further developed in [18]. We use their code which produces "optimized LHS" via simulated annealing.

We also used "maxmin LHS" obtained by selecting an LHS design from a set of four different designs according to the maxmin criterion. The number of chosen designs was limited by the CPU time and a requirement to use a very large number of sampled points (up to $N = 2^{20}$ averaged over 50 independent replications, see Section 9).

## 4. Quasi Monte Carlo

LDS are specifically designed to place sample points as uniformly as possible. LDS are also known as quasi random numbers. The QMC algorithm for the evaluation of the integral (2.1) has a form similar to (2.2)

$$I_N = \frac{1}{N}\sum_{i=1}^{N} f(q_i). \tag{4.1}$$

Here $\{q_i\}$ is a set of LDS points uniformly distributed in a unit hypercube $H^n$, $q_i = (q_i^1, ..., q_i^n)$.

The Koksma-Hlawka inequality [19] gives an upper bound for the QMC integration error:

$$\varepsilon \leq V(f) D_N^*. \tag{4.2}$$

Here, $V(f)$ is the variation of $f(x)$ in the sense of Hardy and Krause, $D_N^*$ is the sample discrepancy (its definition is given in Section 7). For a one-dimensional function with a continuous first derivative it is simply

$$V(f) = \int_H |df(x)/dx| dx.$$

In higher dimensions, the Hardy-Krause variation may be defined in terms of the integral of partial derivatives. Further it is assumed that $f(x)$ is a function of bounded variation. The smaller the discrepancy $D_N$, the better the convergence of the QMC integration. Numerical experiments suggest that the QMC integration error is determined by the variance of the integrand and not by its variation [20].

It is generally accepted that the rate of the discrepancy – meaning by this the rate at which discrepancy converges as a function of $N$, determines the expected rate of the accuracy. In fact one can use the following estimate of the QMC convergence rate (see Section 7 for details):

$$\varepsilon_{QMC} = \frac{O(\ln N)^n}{N}. \tag{4.3}$$

The QMC rate of convergence (4.3) is much faster than that for the MC method (2.3), although it depends on the dimensionality $n$. Consequently, the smaller the value of $n$, the better this estimate. In practice at $n > 1$ the rate of convergence appears to be approximately equal to $O(N^{-\alpha})$, $0 < \alpha \leq 1$. Hence, the QMC method in most cases outperforms MC in terms of convergence.

# 5. Sobol' sequences

There are a few well-known and commonly used LDS. Many practical studies have proven that the Sobol' LDS is in many aspects superior to other LDS. "Preponderance of the experimental evidence amassed to date points to Sobol' sequences as the most effective quasi-Monte Carlo method for application in financial engineering." [21] (see also [22]). For this reason Sobol' sequences were used in this work. Sobol' sequences are digital ($t$; $s$)-sequences in base 2 in the sense of Niederreiter [19].

Sobol' LDS aim to meet three main requirements [23]:

1. Best uniformity of distribution as $N \to \infty$.
2. Good distribution for fairly small initial sets.
3. A very fast computational algorithm.

Points generated by the Sobol' LDS produce a uniform-looking filling of the space, even for rather small numbers of points $N$. Best uniformity of distribution is defined in terms of low discrepancy and additional uniformity properties A and A' [24]:

**Definition 1.** A low-discrepancy sequence is said to satisfy **Property A** if for any binary segment (not an arbitrary subset) of the $n$-dimensional sequence of length $2^n$ there is exactly one point in each $2^n$ hyper-octant that results from subdividing the unit hypercube along each of its length extensions into half.

**Definition 2.** A low-discrepancy sequence is said to satisfy **Property A'** if for any binary segment (not an arbitrary subset) of the $n$-dimensional sequence of length $4^n$ there is exactly one point in each $4^n$ hyper-octant that results from subdividing the unit hypercube along each of its length extensions into four equal parts.

# 6. Comparison of sample distributions generated by different techniques on a unit square

In this section we compare distributions of $N_1 = 4$ (Fig. 1) and $N_2 = 16$ (Fig. 2) points on a unit square ($n=2$) given by five different sampling techniques: MC, standard LHS, maxmin LHS, optimized LHS and Sobol' LDS. This provides a qualitative picture of the uniformity properties of these sampling techniques. In the first case the unit square is divided into $N_1$ and $N_1^2$ squares of measure $1/N_1$ and $1/N_1^2$ respectively. In the second case the unit square is divided into $N_2$ and $N_2^2$ squares of measure $1/N_2$ and $1/N_2^2$ respectively.

Figs. 1, i and 1, j show 2-dimensional distributions of Sobol' points satisfying Property A. One can see that each of the 4 small squares contains exactly one Sobol' point. This is not the case for standard LHS (Fig. 1,c). Projections to both 1-dimensional subspaces also contain 1 point in each of the 4 intervals for standard LHS (Fig. 1, d), for maxmin LHS (Fig. 1, f), for optimised LHS (Fig. 1, h) and for LDS-Sobol' (Fig. 1, j). Random sampling (Fig. 1,b) does not possess either of these properties.

LDS also possess a more general property: all projection of the *n*-dimensional LDS on *s*-dimensional subspaces ($s < n$) form *s*-dimensional LDS [19, 24]. These additional stratification properties of Sobol' LDS result in increased uniformity of Sobol' sampling. The standard LHS sampling does not possess additional stratification properties in higher dimensions. The LHS design can be improved by optimizing it further. However, this optimization being CPU time consuming is only possible for a small number of points and limited number of dimensions. Although maxmin LHS (Figs. 1,e and 1f) visually shows an improvement in a space distribution over standard LHS, sampled points don't satisfy Property A. However, points produced by optimized LHS do satisfy Property A (Figs. 1,g, and 1,h).

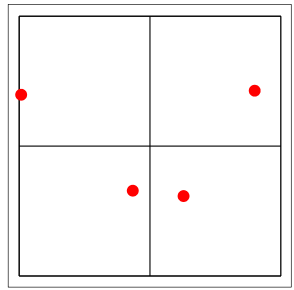
(a)
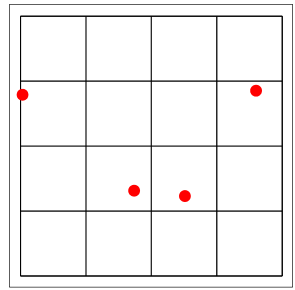
(b)
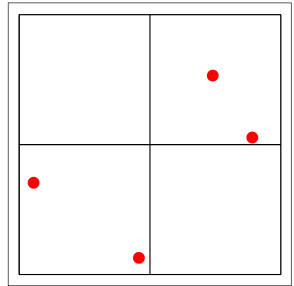
(c)
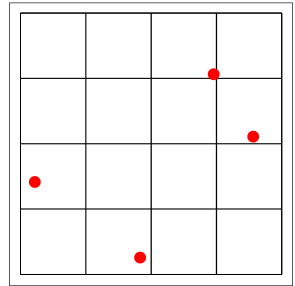
(d)
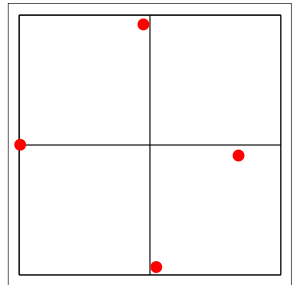
(e)
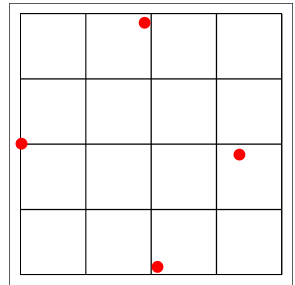
(f)
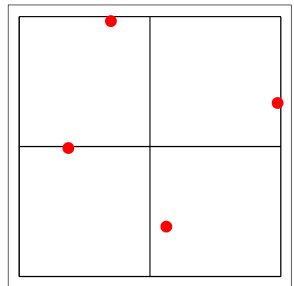
(g)
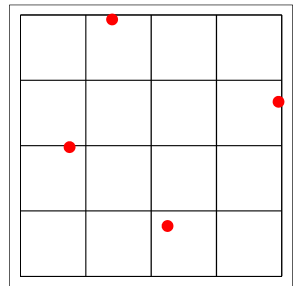
(h)
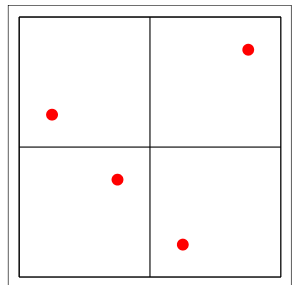
(i)
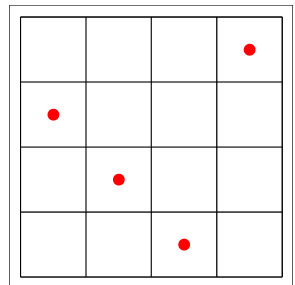
(j)

Fig. 1. Distributions of 4 points in two dimensions. The unit square is divided into 4 (on the left) and 16 (on the right) squares. (a,b) MC, (c,d) standard LHS, (e,f) maxmin LHS, (g,h) optimized LHS, (i,j) Sobol' LDS.

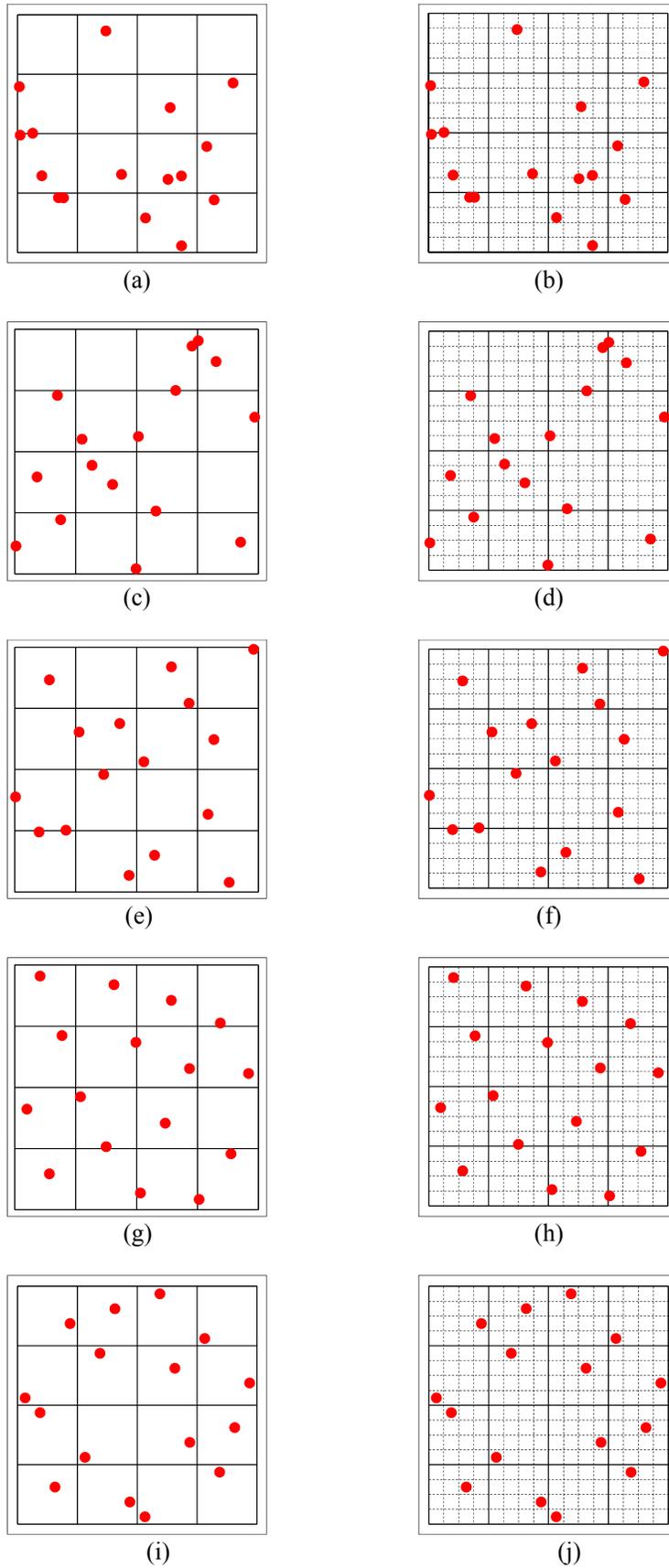

Fig. 2. Distributions of 16 points in two dimensions. The unit square is divided into 16 (on the left) and 256 squares (on the right). (a,b) MC, (c,d) standard LHS, (e,f) maxmin LHS, (g,h) optimized LHS, (i,j) Sobol' LDS.

Fig. 2 shows distributions of 16 points in two dimensions. From Fig. 2, i, and 2, j it is clear that Sobol' points satisfy Property A' in 2 dimensions: each of the $N_2 = 16$ subsquares contains exactly 1 Sobol' point (Fig. 2, i). This is not true for all types of LHS (Fig. 2, c, e, g) and MC (Fig. 2, a) samplings: clustering and empty subsquares are clearly visible from these plots. Optimized LHS gives the best distribution among all other LHS designs. Projections of the 2-dimensional sampling to both 1-dimensional subspaces also contain 1 point in each of the 16 intervals (Figs. 2, j). This is also true for LHS (Fig. 2, d, f, h) but not for MC (Fig. 2, b).

In summary it would appear that Sobol' LDS sampling gives a better way of arranging $N$ points in $n$–dimensions than MC and standard LHS. Although LHS sampling can be improved through optimization, this procedure is limited only for small sample sizes and low dimensions. Another important limitation of optimized LHS is that the sample size cannot be increased incrementally without re-executing the optimization step, which has implications for the monitoring of the convergence of the integration [5].

### 7. Discrepancy

Discrepancy is a quantitative measure for the deviation of sampled points from the uniform distribution. Consider a number of points $N_{Q(t)}$ from a sequence $\{x_i\}$, $i = 1,..,N$ in an $n$-dimensional rectangle $Q(t)$ with the origin in 0, whose sides are parallel to the coordinate axes and which is a subset of $H^n$: $Q(t) \subset H^n$. $Q(t)$ has a volume $m(Q(t)) = t_1 \times ... \times t_n$, where $t = (t_1,...,t_n)$ are the right hand side coordinates of the rectangle with the origin in the centre of coordinates.

The local discrepancy is defined as (see f.e. [19])

$$h(t) = \frac{N_{Q(t)}}{N} - m(Q(t)), \quad (7.1)$$

where $N$ is the total number of points sampled in $H^n$. Fig. 3 illustrates the notion of the local discrepancy in two dimensions.

The star-discrepancy is defined as

$$D_N^* = \sup_{Q(t) \in H^n} |h(t)|. \quad (7.2)$$

By definition a low discrepancy sequence (LDS) is one satisfying the upper bound condition:

$$D_N^* \leq c(n) \frac{(\ln N)^n}{N}. \quad (7.3)$$

Constant $c(n)$ depends upon the sequence as well as upon the dimension, but does not depend on $N$. For random numbers the expected discrepancy is

$$D_N^* = O((\ln \ln N)/N^{1/2}).$$

From the Koksma-Hlawka inequality (4.2) it follows that asymptotically at $N \to \infty$ QMC has a convergence rate $O(1/N)$ which is much better than the rate $O(1/\sqrt{N})$ that can be achieved by MC and LHS. However, this rate can be attained only at the number of sampled points higher than the threshold $N^* \sim \exp(n)$ which is practically not possible at large $n$. The value of discrepancy depends equally on all dimensions. A crucial consideration here is that in practical applications, when one attempts to calculate the quadrature of a given function, some variables (dimensions) are more important than others for any given function. This explains why even for high dimensional problems with hundreds of variables and a number of sampled points $N < N^*$ still QMC has a better convergence rate than MC. The theory behind this behavior is based on global sensitivity analysis and is presented in the next section.

Calculation of the star-discrepancy $D_N^*$ is very difficult and the $L_2$ discrepancy is used instead for practical purposes. The $L_2$ discrepancy is defined as

$$D_N^{L2} = \left(\int_{H^n} h^2(t) dt\right)^{1/2}. \tag{7.4}$$

$D_N^{L2}$ has a closed form [25]:

$$D_N^{L2} = \frac{1}{N^2} \sum_{i,j=1}^{N} \prod_{k=1}^{n}(1-\max(x_i^k - x_j^k)) - \frac{2^{1-n}}{N} \sum_{i=1}^{N} \prod_{k=1}^{n}(1-(x_i^k)^2) + 3^{-n}, \tag{7.5}$$

which was used in this work.

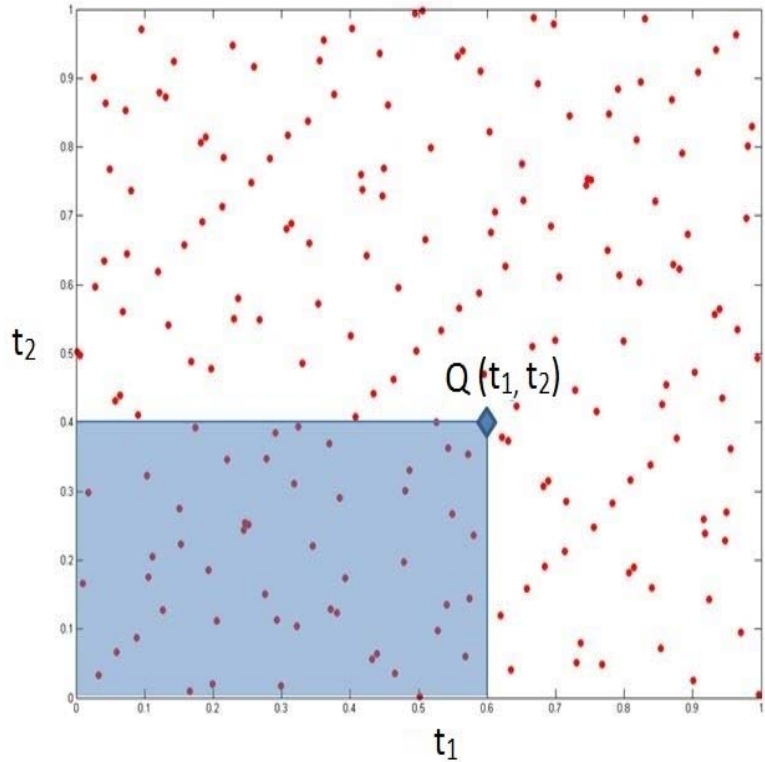

Fig. 3. Illustration of local discrepancy $h(t)$ in two dimensions with Sobol LDS sampling of $N = 256$ points. $N_Q$ is a number of points in the shaded rectangular.

There are also other definitions of discrepancies. Hickernell [26] pointed out that the discrepancy (7.2) is anchored to the origin because the interval [0; *t*) appears in its definition and also there is reference to *t* = (1; … ; 1) since it appears in the formula for the variation in (4.2). He defined the so-called centered discrepancy and variation that refer to the center of the hypercube at *t* = (0.5,…, 0.5). This discrepancy and variation are invariant under reflections of a set of points $N$ about any plane $t_j = 0.5$. The centered discrepancy has some advantages over the standard $L_2$ discrepancy (7.4) when the objective is to identify differences in various sampling designs. Hickernell also introduced the wrap-around discrepancy which measures uniformity not only over the *n*-dimensional unit hypercube $H^n$ but also over all the projection of a set of points $N$ to *s*-dimensional subspaces of $H^s$, $1 \leq s < n$. These results were further developed by Fang *et al* [11] who applied searching algorithms by minimizing discrepancy as the objective function for optimizing LHS designs. Fang *et al* [10] showed that LHS has a lower expectation of square centered discrepancy than that of simple random designs.

We compared uniformity properties of MC, LHS and Sobol' sequences for different dimensions using $D_N^{L2}$ defined by (7.5). Results presented in Fig. 4 were obtained with a standard random number generator as provided by MATLAB®, the LHS generator developed in [17] and the SobolSeq8192 generator [27,28]. The maximum number of sampled points used for calculations was 32768.

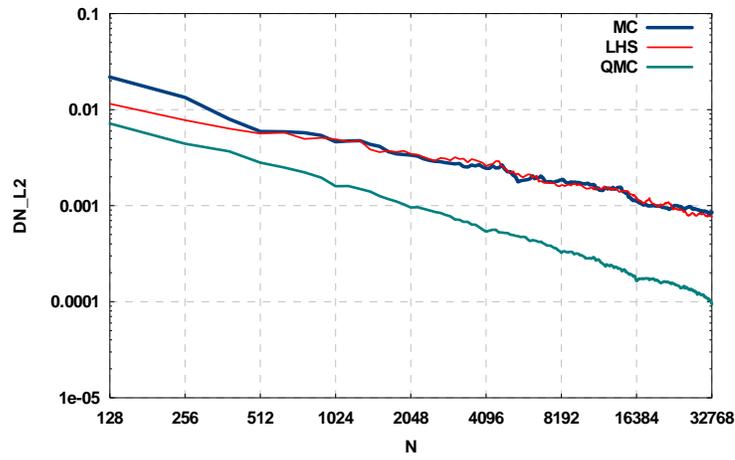

(a)

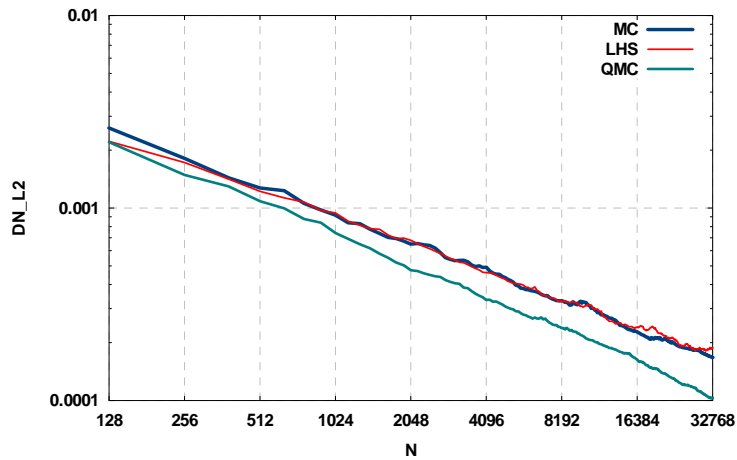

(b)

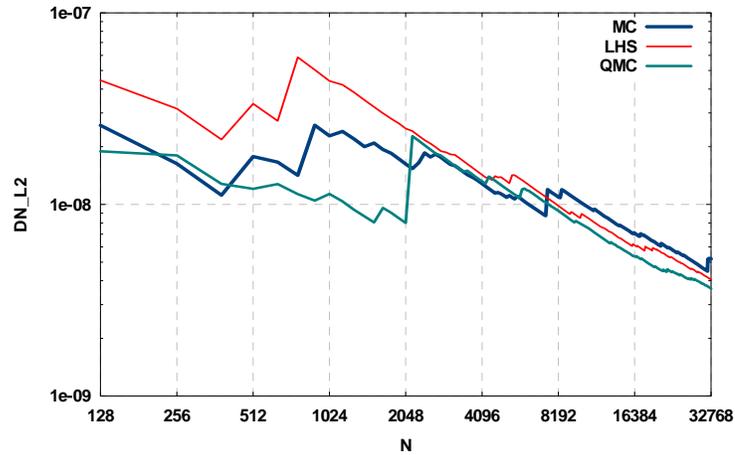

(c)

Fig. 4. $D_N^{L2}$ discrepancy in dimensions $n = 5$ (a), $n = 10$ (b), $n = 40$ (c) versus $N$ for MC (blue broken line), LHS (thick solid red line) and QMC (thin solid green line) samplings.

It can be seen that for low dimensions ($n < 20$) LDS is superior to MC and LHS as it shows the lowest values for the discrepancy. In higher dimensions ($n > 40$) the $L_2$ discrepancy of Sobol' LDS becomes comparable to that of MC and LHS. However as anticipated, discrepancy values are not sufficient to judge the performance of the considered sampling techniques when applied to practical problems such as high dimensional integration. The performance is determined by the effective dimension values for a problem at hand. The notion of the effective dimension is introduced in the next section.

Due to the large number of required sampled points and high dimensions it was not possible to consider optimized LHS. We also considered maxmin LHS using procedure described above. Comparison showed that the results obtained with standard and maxmin LHS are similar, therefore we present only the result of standard LHS – also because standard LHS is the method most seen in the litarature. These findings also concern the results for function integrations presented in Section 9.

It is quite possible that optimized LHS would show a lower $D_N^{L2}$ discrepancy than that of MC. However, optimized LHS can only be used at very low number of sample points $N$ and relatively low dimensions, while all these tests required a very high number of $N$.

## 8. Global Sensitivity Analysis and Effective dimensions

Global Sensitivity Analysis (SA) provides information on the general structure of a function by quantifying the variation in the output variables in response to variation in the input variables.

Consider an integrable function $f(x)$ defined in the unit hypercube $H^n$. It can be expanded in the following form:

$$f(x) = f_0 + \sum_{s=1}^{n} \sum_{i_1 < ... < i_s} f_{i_1...i_s}(x_{i_1},...,x_{i_s}). \tag{8.1}$$

Expansion (8.1) is a sum of $2^n$ components. It can also be presented as

$$f(x) = f_0 + \sum_i f_i(x_i) + \sum_{i<j} f_{ij}(x_i,x_j) + ... + f_{12...n}(x_1,x_2,...,x_n).$$

Each of the components $f_{i_1...i_s}(x_{i_1},...,x_{i_s})$ is a function of a unique subset of variables from $x$. The components $f_i(x_i)$ are called first order terms, $f_{ij}(x_i,x_j)$ - second order terms and so on.

It can be proven [29] that the expansion (8.1) is unique if

$$\int_{H^n} f_{i_1...i_s}(x_{i_1},...,x_{i_s})dx_{i_k} = 0, \ 1 \le k \le s, \tag{8.2}$$

in which case it is called a decomposition into summands of different dimensions. This decomposition is also known as the ANOVA (ANalysis Of VAriances) decomposition. The ANOVA decomposition is orthogonal, *i.e.* for any two subsets $u \ne v$ an inner product

$$\int_{H^n} f_u(x) f_v(x) dx = 0.$$

It follows from (8.1) and (8.2) that

$$\int_{H^n} f(x)dx_1...dx_n = f_0,$$

$$\int_{H^n} f(x) \prod_{k \neq i} dx_k = f_0 + f_i(x_i), \tag{8.3}$$

$$\int_{H^n} f(x) \prod_{k \neq i,j} dx_k = f_0 + f_i(x_i) + f_j(x_j) + f_{i,j}(x_i, x_j)$$

and so on.

For square integrable functions, the variances of the terms in the ANOVA decomposition add up to the total variance of the function

$$\sigma^2 = \sum_{s=1}^{n} \sum_{i_1 < \cdots < i_s} \sigma^2_{i_1 \ldots i_s}, \tag{8.4}$$

where $\sigma^2_{i_1 \ldots i_s} = \int_{H^n} f^2_{i_1 \ldots i_s}(x_{i_1}, \ldots, x_{i_s}) dx_{i_1}, \ldots, x_{i_s}$.

Sobol' defined the global sensitivity indices as the ratios

$$S_{i_1 \ldots i_s} = \sigma^2_{i_1 \ldots i_s} / \sigma^2.$$

All $S_{i_1 \ldots i_s}$ are non negative and add up to one

$$\sum_{s=1}^{n} \sum_{i_1 < \ldots < i_s} S_{i_1 \ldots i_s} = 1.$$

$S_{i_1 \ldots i_s}$ can be viewed as a natural sensitivity measure of a set of variables $x_{i_1}, \ldots, x_{i_s}$. It corresponds to a fraction of the total variance given by $f_{i_1 \ldots i_s}(x_{i_1}, \ldots, x_{i_s})$.

Sobol' also introduced sensitivity indices for subsets of variables [29]. Consider two complementary subsets of variables y and z:

$$x = (y, z).$$

Let $y = (x_{i_1}, \ldots, x_{i_m})$, $1 \leq i_1 < \ldots < i_m \leq n$, $K = (i_1, \ldots, i_m)$. The variance corresponding to $y$ is defined as

$$\sigma^2_y = \sum_{s=1}^{m} \sum_{(i_1 < \cdots < i_s) \in K} \sigma^2_{i_1 \ldots i_s}.$$

$\sigma^2_y$ includes all partial variances $\sigma^2_{i_1}, \sigma^2_{i_2}, \ldots, \sigma^2_{i_1 \ldots i_s}$ such that their subsets of indices $(i_1, \ldots, i_s) \in K$.

The total variance $(\sigma^{tot}_y)^2$ is defined as

$$(\sigma^{tot}_y)^2 = \sigma^2 - \sigma^2_z$$

$(\sigma^{tot}_y)^2$ consists of all $\sigma^2_{i_1 \ldots i_s}$ such that at least one index $i_p \in K$ while the remaining indices can belong to the complementary to $K$ set $\bar{K}$ [29]. The corresponding global sensitivity indices are defined as

$$S_y = \sigma^2_y / \sigma^2,$$

$$S^{tot}_y = (\sigma^{tot}_y)^2 / \sigma^2.$$

$S_y^{tot} = 1 - S_z$, $S_y^{tot} - S_y$ accounts for all interactions between *y* and *z*.

The important indices in practice are $S_i$ and $S_i^{tot}$ [31,32]. Sampling based computational strategies to compute these measures are discussed in [33,34]. The first order index, written as

$$S_i = \frac{V_{x_i}\left(E_{x_{\sim i}}\left(f(x)|x_i\right)\right)}{V(f(x))}$$

is identical to Pearson's correlation ratio [35]. Here $x_{\sim i}$ denotes a set of variables with indices different from index *i*.

In most cases knowledge of $S_i$ and $S_i^{tot}$ provides sufficient information to determine the sensitivity of the function being investigated to individual input variables. Their values also can be used to determine function effective dimensions. The notion of the "effective dimension" was introduced in [36]. Let |y| be a cardinality of a subset $y$.

**Definition.** The effective dimension of $f(x)$ in the superposition sense is the smallest integer $d_S$ such that

$$\sum_{0<|y|<d_s} S_y \geq 0.99 . \qquad (8.5)$$

**Definition.** The effective dimension in the truncation sense $d_T$ is the smallest integer defined as

$$\sum_{y \subseteq \{1,2,\ldots,d_T\}} S_y \geq 0.99 . \qquad (8.6)$$

The threshold 0.99 is arbitrary. Condition (8.5) means that the function $f(x)$ is almost a sum of $d_S$ – dimensional functions. A small value of $d_S$ implies that there are no high-order interactions. A low value of $d_T$ implies that there are few important variables. The identification of important and not important variables allows non important variables to be fixed at any arbitrary value within their domain of existence. The resultant model has thus lower complexity with dimensionality reduced from $n$ to $d_T$. Condition $d_T \ll n$ is often verified in practical problems.

The value $d_S$ does not depend on the order in which the input variables are sampled, while $d_T$ does. The following inequality is always satisfied: $d_S \leq d_T$. For some problems changing the order in which input variables are sampled can dramatically decrease $d_T$ (see *e.g.* [36], [37], [38]).

A straightforward evaluation of the effective dimensions from their definitions is not practical in general. Global sensitivity analysis allows to estimate the effective dimensions at reasonable computational costs as discussed in [14].

## 9. Results. Integration

It this section we consider the integral evaluation for different classes of functions, reporting the integration error and the rate of convergence. A taxonomy for functions' difficulty was suggested in [14] and is based on the functions' effective dimensions:

*Type A*: Functions with not equally important variables. For type A models the effective dimension in the truncation sense $d_T \ll n$.

*Type B*: Functions with equally or almost equally important variables and with dominant low order interaction terms in ANOVA decomposition. For such models the effective dimension in the superposition sense $d_S \ll n$, while $d_T \approx n$.

*Type C*: Functions with equally or almost equally important variables and with dominant high order interaction terms in ANOVA decomposition. For type C models $d_S \approx d_T \approx n$.

This classification is summarized in Table 1. Due to the superimposition of high cardinality and high interaction class C represents the most difficult class of functions for integrals evaluation.

Information about relative efficiencies of the three sampling techniques is presented in the last two columns.

Table 1 Classification of functions based on the effective dimensions. Two complementary subsets of variables y and z are considered: $x = (y, z)$.

| Function type | Description | Relationship between $S_i$ and $S_i^{tot}$ | $d_T$ | $d_S$ | QMC is more efficient than MC | LHS is more efficient than MC |
|---|---|---|---|---|---|---|
| A | A few dominant variables | $S_y^{tot}/n_y >> S_z^{tot}/n_z$ | $<< n$ | $<< n$ | Yes | No |
| B | No unimportant subsets; only low-order interaction terms are present | $S_i \approx S_j, \forall\, i, j$ $S_i / S_i^{tot} \approx 1, \forall\, i$ | $\approx n$ | $<< n$ | Yes | Yes |
| C | No unimportant subsets; high-order interaction terms are present | $S_i \approx S_j, \forall\, i, j$ $S_i / S_i^{tot} << 1, \forall\, i$ | $\approx n$ | $\approx n$ | No | No |

For types A and B functions QMC integration can attain the rate of convergence close to the theoretical limit $O(1/N)$ regardless of the nominal dimension *n*, although the presence of high order interaction terms in the ANOVA decomposition can somewhat decrease the convergence rate.

The ANOVA decomposition in a general case can be presented as $f(x) = f_0 + \sum_i f_i(x_i) + r(x)$, where $r(x)$ are the ANOVA terms corresponding to higher order interactions. Stein [6] showed that variance computed with the LHS design is

$$E(\varepsilon_{LHS}^2) = \frac{1}{N} \int_{H^n} [r(x)]^2 dx + O(\frac{1}{N}),$$

while for MC

$$E(\varepsilon_{MC}^2) = \frac{1}{N} \sum_i \int_{H^n} [f_i(x_i)]^2 dx + \frac{1}{N} \int_{H^n} [r(x)]^2 dx + O(\frac{1}{N}).$$

In the ANOVA decomposition of type B functions, the effective dimension $d_S$ is small, hence $r(x)$ is also small comparing to the main effects. In the extreme case it is equal to 1, and a function $f(x)$ can be presented as a sum of one-dimensional functions $f(x) = f_0 + \sum_i f_i(x_i)$. This means that only one-dimensional projections of the sampled points play a role in the function approximation. For type B functions LHS can achieve a much higher convergence rate than that of the standard MC.

To test the classification presented above MC, LHS and QMC integration methods were compared considering test functions presented in Tables 2-4. All functions are defined in $H^n$. The theoretical values of all integrals apart from the function 1A are equal to 1. For the function 1A a value of an integral is equal to $I[f]=-\frac{1}{3}(1-(-\frac{1}{2})^n)$.

Table 2 Convergence results for Type A functions

| Index | Function $f(x)$ | Dim $n$ | Slope $\alpha$ MC | Slope $\alpha$ QMC | Slope $\alpha$ LHS |
|---|---|---|---|---|---|
| 1A | $\sum_{i=1}^{n}(-1)^i \prod_{j=1}^{i} x_j$ | 360 | 0.50 | 0.94 | 0.52 |
| 2A | $\prod_{i=1}^{n} \frac{|4x_i - 2| + a_i}{1 + a_i}$<br><br>$a_1 = a_2 = 0$<br>$a_3 = \ldots = a_{100} = 6.52$ | 100 | 0.49 | 0.65 | 0.50 |

Table 3 Convergence results for Type B functions

| Index | Function $f(x)$ | Dim $n$ | Slope $\alpha$ MC | Slope $\alpha$ QMC | Slope $\alpha$ LHS |
|---|---|---|---|---|---|
| 1B | $\prod_{i=1}^{n} \frac{n - x_i}{n - 0.5}$ | 30 | 0.52 | 0.96 | 0.69 |
| 1B | $\prod_{i=1}^{n} \frac{n - x_i}{n - 0.5}$ | 100 | 0.50 | 0.94 | 0.75 |
| 2B | $\left(1+\frac{1}{n}\right)^n \prod_{i=1}^{n} \sqrt[n]{x_i}$ | 30 | 0.50 | 0.87 | 0.62 |
| 2B | $\left(1+\frac{1}{n}\right)^n \prod_{i=1}^{n} \sqrt[n]{x_i}$ | 100 | 0.49 | 0.79 | 0.71 |

Table 4 Convergence results for Type C functions

| Index | Function $f(x)$ | Dim $n$ | Slope $\alpha$ MC | Slope $\alpha$ QMC | Slope $\alpha$ LHS |
|---|---|---|---|---|---|
| 1C | $\prod_{i=1}^{n}|4x_i - 2|$ | 10 | 0.47 | 0.64 | 0.50 |
| 2C | $(1/2)^{1/n} \prod_{i=1}^{n} x_i$ | 10 | 0.49 | 0.68 | 0.51 |

Figs. 5-8 show the root mean square error (RMSE) versus the number of sampled points. The root mean square error is defined as

$$\varepsilon = \left( \frac{1}{K} \sum_{k=1}^{K} (I[f] - I_N^k[f])^2 \right)^{1/2},$$

where $K$ is a number of independent runs, $I_N^k[f]$ is a value of the MC/QMC estimate of $I[f]$ with the use of $N$ sampled points at $k$-th independent run. For the MC and LHS method all runs were statistically independent. For QMC integration for each run a different part of the Sobol' LDS was used. For all tests $K=50$. The RMSE is approximated by the formula $cN^{-\alpha}$, $0<\alpha<1$. Smaller $c$ means smaller RMSE. The value of $\alpha$ defines the convergence rate. The trend lines and corresponding values for $\alpha$ in parenthesis are presented in Figures 5-8 and in the last three columns of Tables 2-4.

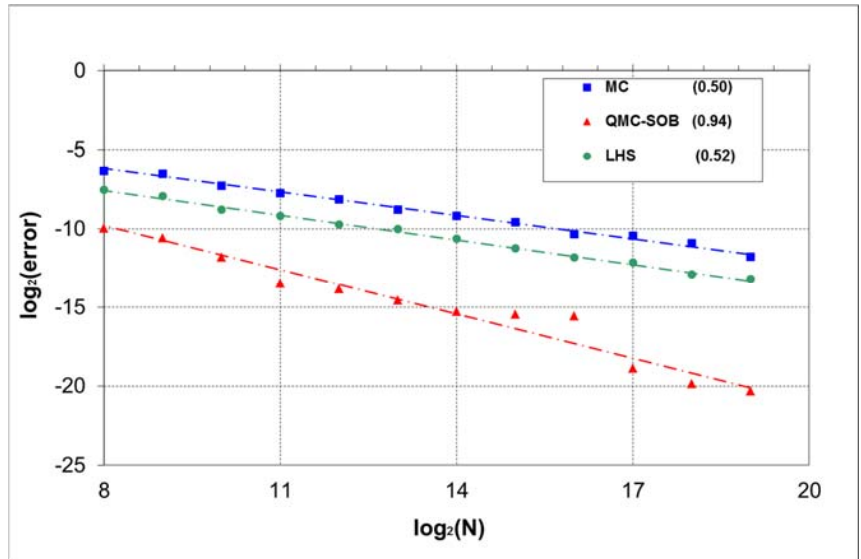

(a)

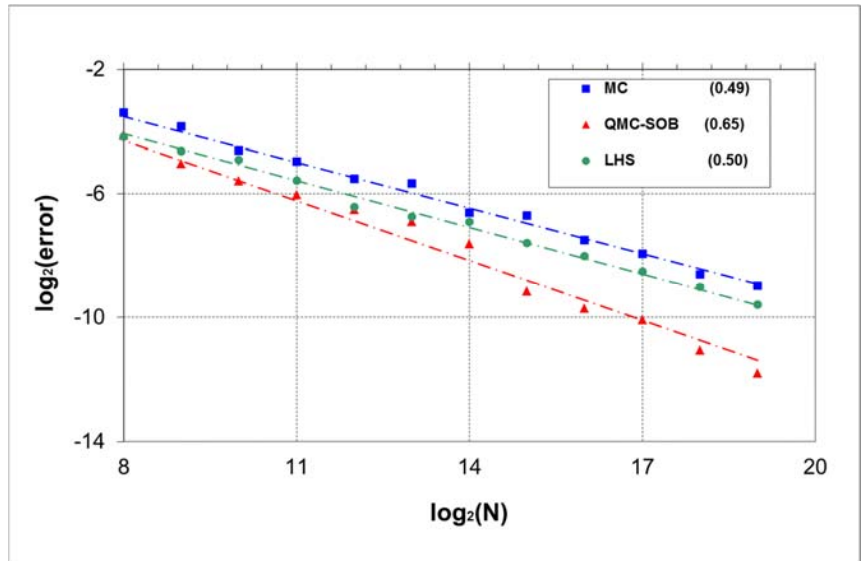

(b)

Fig. 5. RMSE versus the number of sampled points for type A models.
(a) function 1A ($n = 360$); (b) function 2A ($n = 100$).

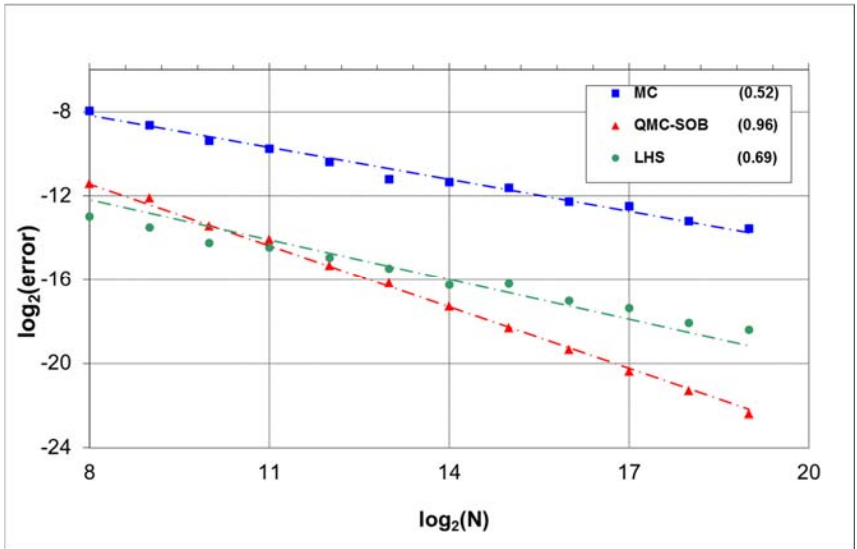
(a)

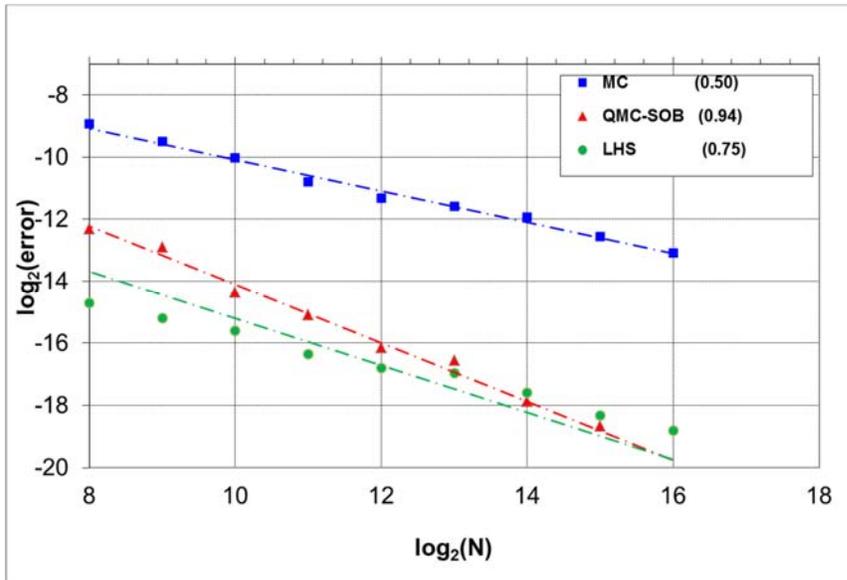
(b)

Fig. 6. RMSE versus the number of sampled points for type B models, function 1B.
(a) function 1B ( $n = 30$ ); (b) function 1B ( $n = 100$ ).

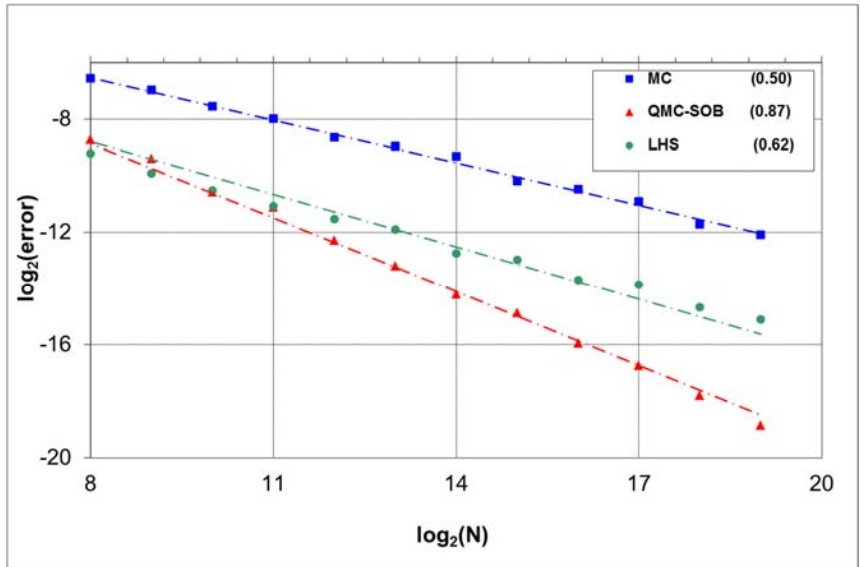

(a)

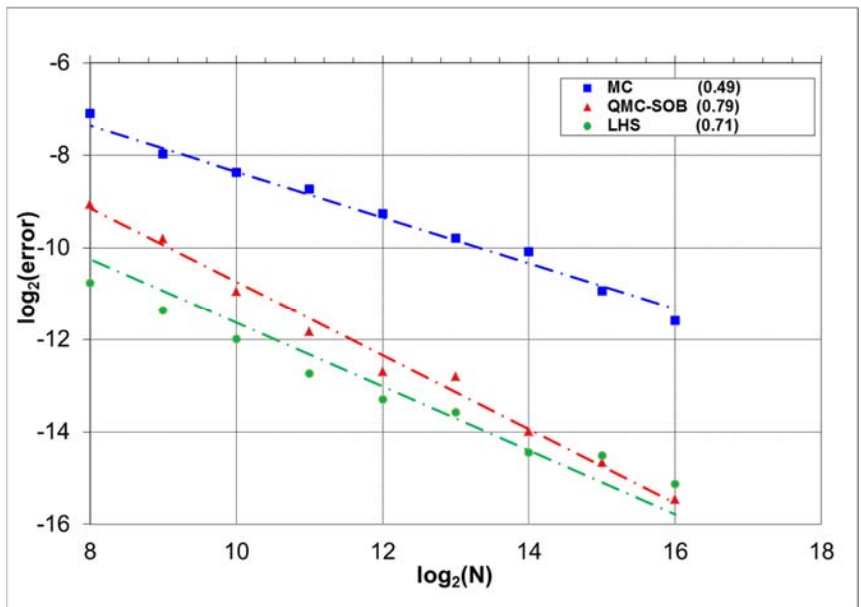

(b)

Fig. 7. RMSE versus the number of sampled points for type B models, function 2B.
(a) function 2B ( $n = 30$ ); (b) function 2B ( $n = 100$ )

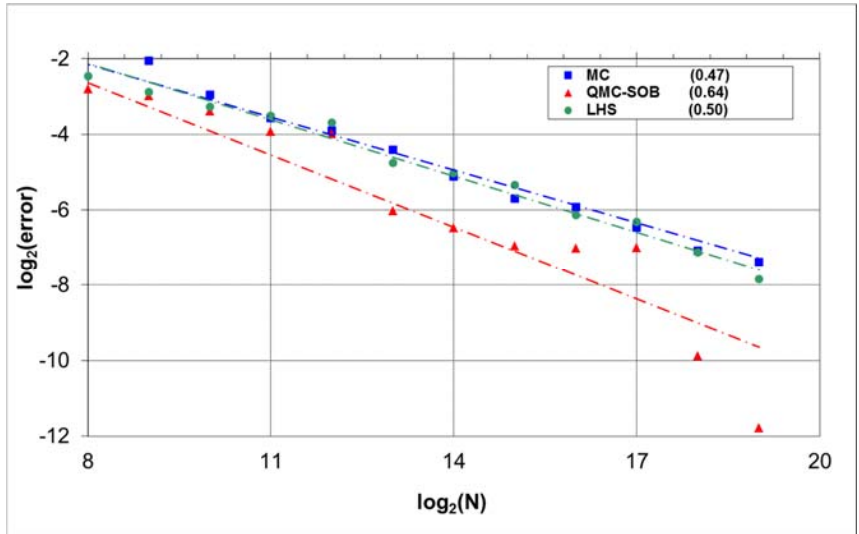
(a)

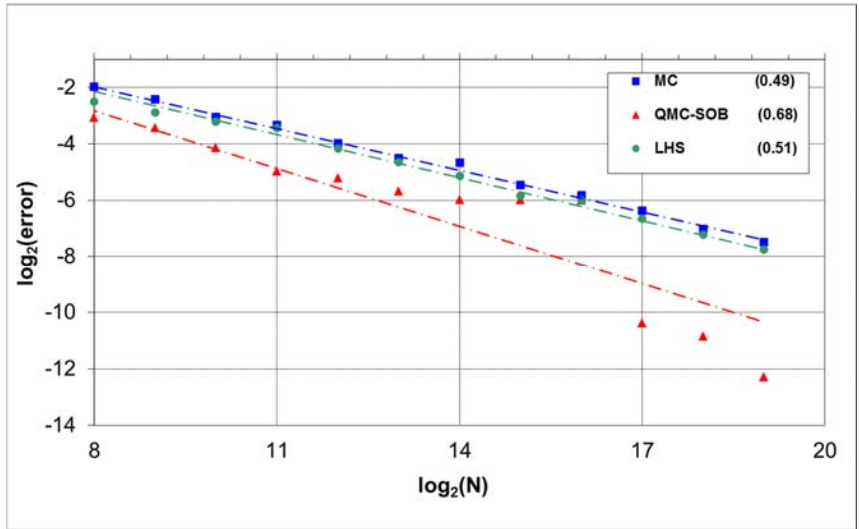
(b)
Fig. 8. RMSE versus the number of sampled points for type C models.
(a) function 1C ( $n = 10$ ); (b) function 2C ( $n = 10$ )

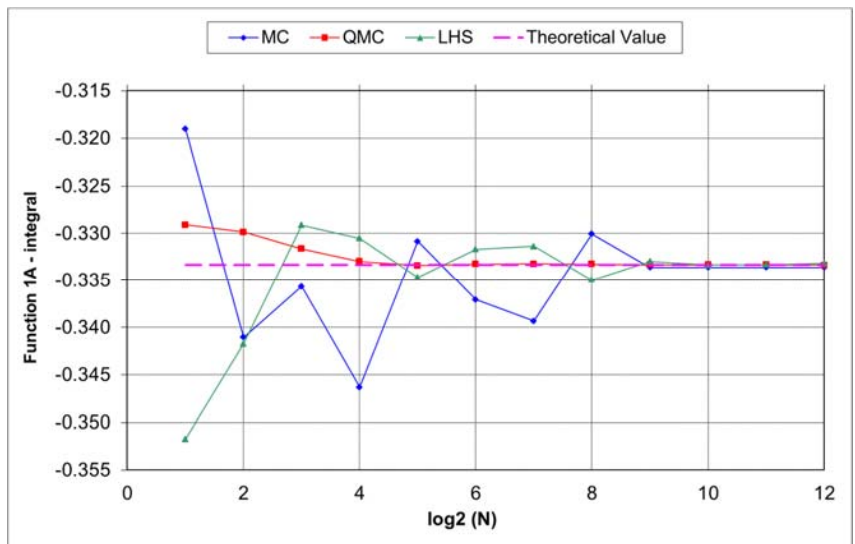

Fig. 9. Values of an integral for Function 1A versus $N$ for MC, QMC and LHS sampling methods ($n = 360$).

These results show that in agreement with theory, QMC integration is superior to that of MC and LHS both in terms of the rate of convergence (larger $\alpha$) and in the absolute value of the integration error (constant $c$) in all test cases.

For function 1A at $n=360$ (a very high dimensional problem) the exponent for algebraic decay in the case of QMC integration $\alpha_{QMC} = 0.94$ is very close to the theoretically predicted asymptotical value $\alpha = 1$. The constant $c$ is lower for the QMC method than that for both MC and LHS methods. The LHS method shows the same convergence rate $\alpha \approx 0.5$ as the MC method and it has only a marginally smaller constant $c$ than that of MC.

For function 2A at $n = 100$ the efficiency of QMC is higher than that of both MC and LHS, although $\alpha_{QMC}$ is only close to 0.7. As for the previous case efficiencies of MC and LHS methods are similar with LHS having a marginally smaller constant $c$.

Integrals for type B functions are considered at two different dimensions ($n = 30$ and $n = 100$) to show an interesting effect of interplay between the efficiencies of QMC and LHS methods. At $n = 30$ $0.87 \leq \alpha_{QMC} \leq 0.96$, while $0.62 \leq \alpha_{LHS} < 0.69$ and $\alpha_{MC} \approx 0.5$. The constant $c$ is lower for the QMC method than that for the MC method at all $N$ and the LHS method at $N > 2^{10}$. At $n = 100$ the efficiency of the QMC method slightly drops, with $0.79 \leq \alpha_{QMC} \leq 0.94$, while the efficiency of the LHS method slightly increases with $0.71 \leq \alpha_{LHS} \leq 0.75$. The efficiency of the MC method remains unchanged: $\alpha_{MC} \approx 0.5$. The constant $c$ is lower for the LHS method at $N \square\ 2^{14} \div 2^{15}$ than that for the QMC method, while at higher $N > 2^{15}$ due to the higher gradient the QMC method becomes more efficient than the LHS method.. We note, that type B is the only type of functions for which LHS is significantly more efficient than MC.

For the most difficult type C functions we presented results only for a relatively low dimensional case of $n = 10$. Here the QMC method remains the most efficient method, while LHS shows practically no better performance than MC. However, in higher dimensions the convergence rate of QMC integration drops and becomes close to that of the other two sampling methods.

Fig. 9 presents convergence plots for the function 1A without averaging over independent runs. For the QMC method the convergence is monotonic and very fast. A convergence curve reaches the theoretical value at $N \approx 2^4$, while for both MC and LHS convergence curves are oscillating and approach the theoretical value only at $N \approx 2^9$. The QMC method is approximately 30 times faster than the LHS method.

The advantage of using MC or QMC schemes over LHS is that new points can be added incrementally: formulas (2.2) and (4.1) can be written as

$$I_N = \frac{N-1}{N} I_{N-1} + \frac{1}{N} f(x_N).$$

Here $f(x_N)$ is the new updated value of an integrand (for the QMC method its value is equal to $f(q_N)$). Thus when using QMC it is straightforward to code a termination criterion that can be invoked incrementally. For LHS instead it is not possible to incrementally add a new point while keeping the old LHS design: in formula (3.1) random permutations of $\{1,..., N-1\}$ are different from random permutations of $\{1,..., N\}$. This means that adding a new $N$th point to an already sampled set of $N$ points and a set $\{f(x_i)\}$, $i = 1,..., N-1$ of function values requires to resample all $N$ points and to recalculate the whole set $\{f(x_i)\}$, $i = 1,..., N-1$ of function values at new $\{x_i\}$, $i = 1,..., N-1$.

## 10. Results. Evaluation of quantiles for normally distributed variates

Many practical simulation problems require the generation of normally distributed variables. In this section we compare MC, LHS and QMC sampling methods for evaluation of quantiles. We consider evaluation of low 5% and high 95% percentiles for the cumulative distribution function of a random variable $f(x) = \sum_{i=1}^{n} x_i^2$, where $x_i$ are independent standard normal variates, $n$ is the dimension, $n = 5$. The considered function is distributed according to the chi-squared distribution with 5 degrees of freedom: $f(x) \sim \chi^2(5)$. The values of 5% and high 95% percentiles are 1.146 and 11.071, respectively.

The algorithm for generating $n$-dimensional random normal variables using uniformly distributed variables consists of the following steps:

a) generate $n$-dimensional random vector $u$ uniformly distributed between 0 and 1 using random numbers, LHS or quasi random sequences;

b) transform every element of $u_i$ into a standard normal vector $\tilde{x}_i$ with zero mean and unit variance using the inverse normal cumulative distribution function: $\tilde{x}_i = F^{-1}(u_i)$.

Fig. 10 presents convergence of numerical estimates to the true values of quantiles.

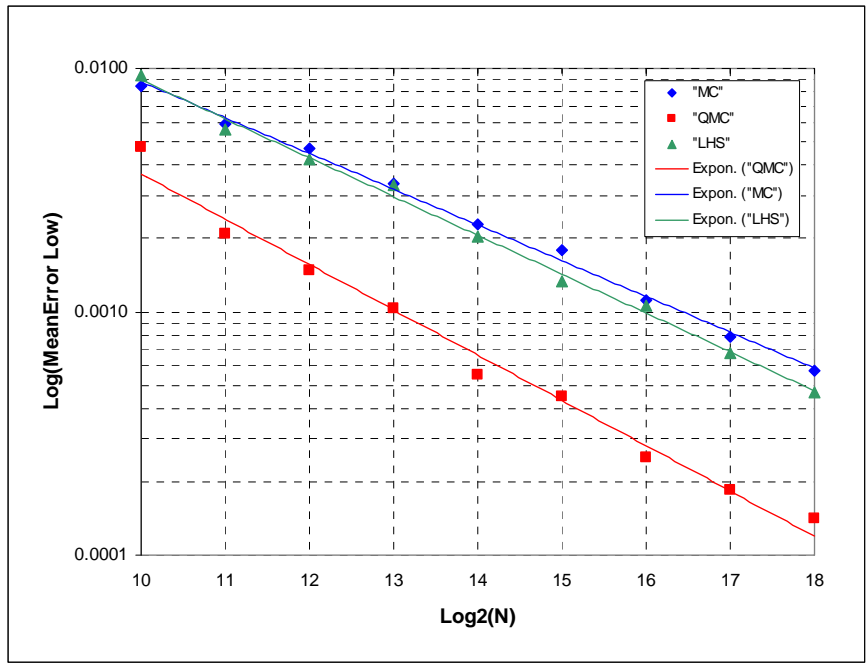

(a)

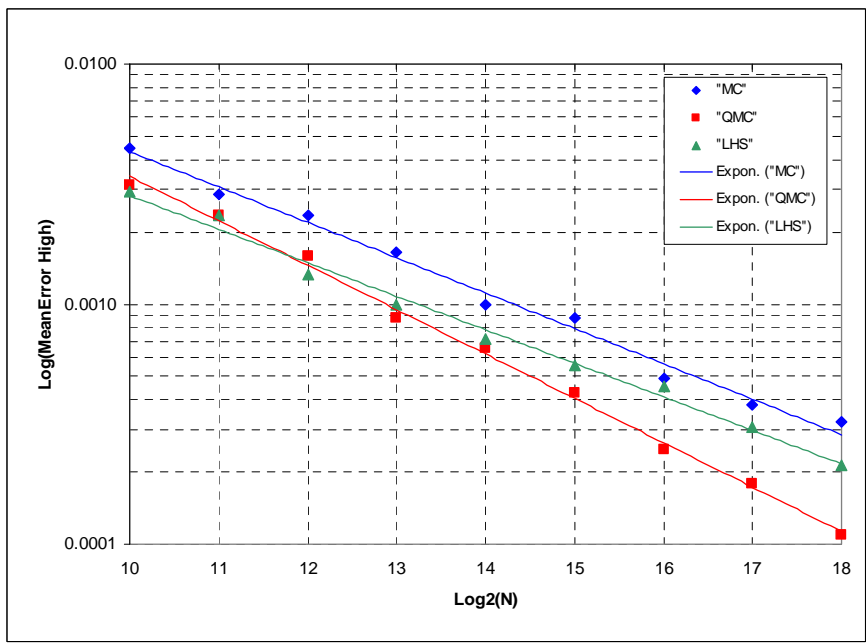

(b)

Fig. 10. Comparison between MC, LHS and QMC methods for evaluation of quantiles: RMSE error versus $N$. (a) Low quantile = 0.05; (b) High quantile = 0.95, dimension $n$ = 5. MC – diamonds, LHS – triangles, QMC – squares.

Results show a superior convergence of the QMC method. LHS provides marginally higher convergence than MC.

**11. Conclusions**

We compared the efficiencies of three sampling methods: MC, LHS and QMC based on Sobol' sequences. It was shown that Sobol' sequences possess uniformity properties which other sampling techniques do not have (Properties A and A'). Optimized LHS satisfies Property A in low dimensions but in high dimensions due to the cost high of optimization it is very difficult to verify this property. In any case while for optimized LHS the fulfilment of the property has to be estimated empirically case by case, Sobol' sequences possess these additional uniformity properties by construction. Comparison of $L_2$ discrepancies revealed that the QMC method has the lowest discrepancy up to dimension 20.

We used a number of test functions of various complexities for high dimensional integration. Comparison showed that for types A and C functions LHS shows only a slight improvement over MC. The convergence rate of the QMC method for types A and B functions is close to $O(1/N)$, while the MC method has the convergence rate close to $O(1/\sqrt{N})$ and the LHS method outperforming MC only for type B functions. For type B functions the superiority of the QMC method over the LHS method can only be seen at some large number of sampled points, while at small number of sampled points the LHS method can have a smaller variance than the QMC method.

For type C functions convergence of the QMC method significantly drops, however it still remains the most efficient method among the three sampling techniques. We also compared MC, LHS and QMC sampling methods for evaluation of quantiles and showed that QMC remains the most efficient method among the three techniques.

Although there has been a significant progress in improving space filling properties of LHS, due to the high cost of optimization involved, even the most efficient techniques cannot be applied for solving high dimensional problems. On the other side optimized LHS – not tested in the present work - have proven to be very effective in application to metamodelling where often only a small number of sampled points are required and a sampling design of such points can be efficiently optimized.

An important practical disadvantage of LHS over MC or QMC schemes is that new points cannot be added sequentially, which makes it very difficult to use in integration where termination criteria are often invoked incrementally and in other methods such as adaptive sampling.

Note that in most instances the evaluation of integrals need to be obtained for functions of unknown typology (e.g. the function is a computer code). LHS performs better than QMC only for class B functions. These functions – where all variable are in a sense important but the interactions are low – are hardly the most frequent – they need built ad hoc because the importance of factors in mathematical models customarily follow a Pareto law – with few important factors and many non-important ones (Saltelli et al., 2008). It would thus appear that a safe bet in the choice of a sampling algorithm for quadratures is provided by Quasi Random Numbers – and this work has made the case that it is so using Sobol' sequences. The present work may be put in perspective by noting that by far the majority of application seen in the literature use non-optimized LHS.

**Acknowledgements**

The authors would like to thank Bertrand Iooss for his valuable comments and providing codes for producing optimized LHS. The European Centre for Governance in Complexity is a joint undertaking of the Centre for the Study of the Sciences and the Humanities (SVT), University of Bergen (UIB) and of the Institut de Ciència i Tecnologia Ambientals (ICTA), Universitat Autonoma de Barcelona.